\documentclass[aps,pra,twocolumn]{revtex4-1}
\usepackage{graphics}
\usepackage{dcolumn}
\usepackage{bm}
\usepackage{amsmath,epstopdf}%
\usepackage{hyperref}
\hypersetup{colorlinks=true, urlcolor=blue}

\usepackage{newlfont}
\usepackage{epstopdf}
\usepackage{amssymb}
\usepackage{amsfonts}
\usepackage{amsmath}
\usepackage{graphicx}
\usepackage{comment}

\newcommand{\bra}[1]{\left< #1 \right|}
\newcommand{\ket}[1]{\left| #1 \right>}

\DeclareMathOperator{\Tr}{Tr}

\begin{document}

\renewcommand*{\DefineNamedColor}[4]{%
   \textcolor[named]{#2}{\rule{7mm}{7mm}}\quad
  \texttt{#2}\strut\\}

\definecolor{red}{rgb}{1,0,0}

\title{Thermal Fluctuations Enhance Order-from-Disorder of Quantum Correlations in Quenched Disordered Spin Models}
\author{Debasis Sadhukhan, R. Prabhu, Aditi Sen(De), Ujjwal Sen}
\affiliation{Harish-Chandra Research Institute, Chhatnag Road, Jhunsi, Allahabad 211 019, India}


\begin{abstract}
We consider paradigmatic quenched disordered quantum spin models, viz., the $XY$  spin glass and random-field $XY$ models, and show that quenched averaged quantum correlations can exhibit the order-from-disorder phenomenon for finite-size systems as well as in the thermodynamic limit. Moreover, we find that the order-from-disorder can get more pronounced in the presence of temperature by suitable tuning of the system parameters. The effects are found for entanglement measures as well as for information-theoretic quantum correlation ones, although the former show them more prominently. We also observe that the equivalence between the quenched averages and their self-averaged cousins -- for classical and quantum correlations -- is related to the quantum critical point in the corresponding ordered system. 
\end{abstract}


\maketitle

\section{Introduction}

Perfectly ordered systems are hard to prepare in the laboratory due to the presence of several uncontrollable factors and hence disorder appears almost inevitably in most systems. The presence of impurities, dislocations of atoms from their regular lattice sites, and environmental effects on the system lead to disorder. Defects can also be modeled, by introducing non-uniform tuning parameters or allowing coupling between random sites \cite{selfavebook,Lewenstein,Sachdev,Misguich, Ahufinger,somen}. Intuitively, one expects that disorder would reduce the properties like magnetization and conductivity of a system, and this is indeed true for a large variety of systems \cite{classical-corr,Niederberger2008,Niederberger2010,Sougato1, Sougato2, Sougato3, Sougato4, Sougato5, Sougato6, Sougato7}. However, there are examples of certain systems, both classical as well as quantum, in which properties like magnetization, classical correlators, entanglement get enhanced with the introduction of disorder -- the phenomena are termed as order-from-disorder \cite{Villain, Wehr, Aharony, Niederberger2008,Niederberger2010,PrabhuODO,UtkarshCI,zanardi}. Moreover, disordered systems, in general, possess rich phases like Bose glass \cite{Boseglass} and spin glass \cite{spinglass} and support phenomena like high $\textrm{T}_{\textrm{c}}$-superconductivity \cite{hightTsupc} and Anderson localization \cite{Alocal}. Recent developments in experimental techniques give rise to the possibility of observing such phenomena in laboratories \cite{DOexpts}. 

Many-body systems can be useful substrates to realize several quantum information protocols \cite{Lewenstein,vedralRMP}. In recent years, the behavior of quantum correlations in many body systems at zero as well as at finite temperatures have been extensively studied \cite{Lewenstein,nonmonotonicity, vedralconf, vedralRMP}. However, most of the studies are restricted to ordered systems \cite{nielsen1}.

In classical systems, phase transitions occur only due to thermal fluctuations \cite{Binneybook}, while quantum systems can have fluctuations even at zero temperature, and may lead to quantum phase transitions \cite{Sachdev}.  Since the absolute zero temperature is inaccessible in the laboratory, characterizing systems at low temperature is important from the point of view of observing the physical properties in experiments. Moreover, at finite temperatures, the interplay between thermal and quantum fluctuations may lead to non-intuitive co-operative phenomena. 

In this paper, we deal with the thermal state of the quenched disordered anisotropic $XY$ spin chain with periodic boundary conditions. Specifically, disorder is introduced either in the coupling constant -- quantum $XY$ spin glass -- or in the strength of the external magnetic field -- random-field quantum $XY$ model. Although disordered $XY$ models cannot be solved analytically like the ordered ones, the single- and two-site properties of the ground and thermal equilibrium states of the disordered $XY$ models can be investigated for reasonably large system sizes using the Jordan-Wigner transformation \cite{LSM,barouch1, barouch2}. To compare the properties between ordered and disordered models, we introduce a quantity called the enhancement score \cite{UtkarshCI} corresponding to any physical observable. This quantity can be used to quantify the order-from-disorder phenomena. We analyze systems of up to $10^3$ quantum spins-$\frac 12$ particles, and find that  entanglement measures \cite{HHHH} like concurrence \cite{conc} and logarithmic negativity \cite{logneg}, and information-theoretic quantum correlation measures \cite{kavanRMP} like quantum discord \cite{discord} and quantum work-deficit \cite{wd}, exhibit a positive enhancement score both at zero and finite temperatures, irrespective of the value of the anisotropy constant. Moreover, we find that there exists a range in the parameter space in which enhancement scores for quantum correlations is higher at finite temperature than that at zero temperature.  Such enhancement due to thermal fluctuations is more pronounced in case of entanglement measures in comparison to that of information-theoretic ones. As a by-product, we show that there are distinct regions in the parameter space, of the post-quenched regime, where self-averaging happens for the quantum correlations, and where the same does not happen, and the regions are related to the quantum critical point of the ordered chain

The paper is organized as follows. In Sec. \ref{sec:model}, we discuss the technique to handle the one-dimensional quantum $XY$ model with transverse field, for both ordered as well as disordered systems. Here, we briefly outline the method to evaluate the correlation functions and magnetizations. In Sec. \ref{sec:EnhancementScore}, we give definitions of the enhancement score to characterize order-from-disorder. The definitions of the quantum correlation measures, used in this paper, are given in Sec. \ref{sec:qcmeasures}. The quenching and self-averaging of observables are discussed in Sec. \ref{subsec:quench}. Sec. \ref{sec:mainresults} presents the  results on order-from-disorder and its nature using the enhancement scores. In particular, the enhancements of quantum correlation measures in the presence of disorder in system parameters and thermal fluctuations are presented. We conclude in Sec. \ref{sec:conclusion}.

\section{The Models and The Method}
\label{sec:model}

We briefly review here the exact diagonalization technique for the $XY$ spin chain by  Jordan-Wigner, Fourier, and Bogoliubov transformations \cite{LSM, barouch1, barouch2}. The method helps us to evaluate physical quantities in the disordered case for relatively large system size \cite{mckenzieprl, mckenzie}. \\

The Hamiltonian for the anisotropic $XY$ model with nearest-neighbor interaction on a one-dimensional (1D) lattice, with $N$ sites in a transverse field is given by 

\begin{equation}
H =   \sum_{i = 1}^{N} \frac {J_i}{4} \Big[(1+\gamma) \sigma_i^x \sigma_{i+1}^{x} + (1 - \gamma ) \sigma_i^y \sigma_{i+1}^{y}\Big] -  \sum_{i=1}^N \frac {h_i}{2} \sigma_i^z,
\label{eqn: XY}
\end{equation}
where $J_i$ is the coupling constant between nearest-neighbor sites $i$ and $i+1$, $h_i$ represents the transverse field strength at the $i^\textrm{th}$ site, and $\gamma\, (\neq 0)$ is the anisotropy constant. 
Here, $\sigma_i^j\, (j=x,y,z)$ corresponds to the Pauli spin matrices at the $i^\textrm{th}$ site. 
In case of the ordered system, we assume all the $J_i$ and $h_i$ are separately equal and  we denote them by $J$ and $h$ respectively. In this paper, we assume periodic boundary conditions, so that $\vec\sigma_{N+1}=\vec\sigma_1$.

The procedure, used to solve Eq. (\ref{eqn: XY}) is to map the Pauli spin operators to spinless fermions via the Jordan-Wigner transformation and thereby, Eq. (\ref{eqn: XY}) reduces to 
%
(neglecting additive constants)
\begin{equation}
H = \sum_{i,j = 1}^{N} c_i^{\dagger}A_{ij}c_j + \frac 12\sum_{i,j=1}^{N} \big( c_i^{\dagger}B_{ij}c_{j+1}^{\dagger} + h.c. \big),
\label{eqn: XYAB}
\end{equation}
where $A$ and $B$ are symmetric and antisymmetric real $N \times N$ matrices, respectively, and are given by 
\begin{align}
A_{ij} &= -h_i\delta_{ij} + \frac{J_i}{2} \delta_{i+1,j} + \frac{J_{j}}{2} \delta_{i,j+1}, \nonumber\\
B_{ij} &= \frac{\gamma}{2} (J_i \delta_{i+1,j} - J_j \delta_{i.j+1}), \nonumber
\end{align}
with $A_{1N}=A_{N1}=J_N$ and  $ B_{1N} = - \frac{\gamma}{2}J_N = - B_{N1}$, to respect the boundary condition. 
 The quadratic Hamiltonian given in Eq. (\ref{eqn: XYAB}) can be diagonalized by using a linear transformation given by
\begin{align}
\eta_k &= \sum_{i=0}^{N-1}\Big( g_{ki}c_i + h_{ki}c_i^{\dagger}\Big), \nonumber \\
\eta_k^{\dagger} &= \sum_{i=0}^{N-1} \Big(g_{ki}c_i{\dagger} + h_{ki}c_i \Big),
\label{eqn: eta}
\end{align}
where $k = -N/2, -N/2+1, \ldots, N/2 - 1$. Here $g_{ki}$ and $h_{ki}$ are real numbers, and the $\eta_{k}$ obey fermionic anticommutation relations. One can express the Hamiltonian in Eq. (\ref{eqn: XYAB}) in terms of the fermionic modes $\eta_{k}$, in such a way that the following two coupled matrix equations hold:
\begin{align}
(A+B)\phi_k^T &=  \Lambda_k\psi_k^T \label{eqn: AA},\\
(A-B)\psi_k^T &= \Lambda_k\phi_k^T \label{eqn: BB}.
\end{align}
Here the components of the two column vectors, $\phi_k^T$  and  $\psi_k^T$, are given by 
\begin{align}
\phi_{ki} &= g_{ki} + h_{ki} \label{eqn: phi},\\
\psi_{ki} &= g_{ki} - h_{ki}  \label{eqn: psi}.
\end{align}
Substituting $\psi_k^T$ from Eq. (\ref{eqn: AA}) to Eq. (\ref{eqn: BB}), we get
\begin{align}
(A-B)(A+B)\phi_k^T = \Lambda_k^2\phi_k^T \label{eqn: EG}.
\end{align}
For $\Lambda_k \ne 0$, one can find $\phi_k^T$ by solving the eigenvalue equation given in Eq. (\ref{eqn: EG}). Then $\psi_k^T$ can be obtained from Eq. (\ref{eqn: BB}). For $\Lambda_k = 0$, both $\phi_k^T$  and  $\psi_k^T$ are determined from Eq. (\ref{eqn: EG}) and their relative signs remains arbitrary. Note here that Eq. (\ref{eqn: EG}) holds for both ordered  and disordered systems.

\subsection{Single- and Two-site Observables for the Ground State}
\label{subsec: methodgs}

At absolute zero, the system freezes to its ground state, $|\Psi_0\rangle$. Let us define two operators, ${\cal A}_i$ and ${\cal B}_i$ in terms of fermionic operators as 
\begin{align}
{\cal A}_i = c_i^{\dagger} + c_i,\,  {\cal B}_i = c_i^{\dagger} - c_i.
\end{align}
The magnetization per site in terms of ${\cal A}_i$ and ${\cal B}_i$ are given by
\begin{align}
m^z_i &=  \bra{\Psi_0}\sigma^z_i\ket{\Psi_0} = -\bra{\Psi_0} {\cal A}_i{\cal B}_i\ket{\Psi_0} \label{eqn: mz},\\
m_i^x &= \bra{\Psi_0}\sigma^x_i\ket{\Psi_0} = \bra{\Psi_0}{\cal A}_i{\cal A}_1{\cal B}_1\cdots {\cal A}_{i-1}{\cal B}_{i-1}\ket{\Psi_0} \label{eqn: mx},\\
m_i^y &=\bra{\Psi_0}\sigma^y_i\ket{\Psi_0} = \bra{\Psi_0}{\cal B}_i{\cal A}_1{\cal B}_1\cdots {\cal A}_{i-1}{\cal B}_{i-1}\ket{\Psi_0}, \label{eqn: my}
\end{align}
by using $\exp{[\pi i c_i^{\dagger}c_i]} = (c_i^{\dagger} + c_i)(c_i^{\dagger} - c_i)$.
The nearest-neighbor diagonal correlation functions take the form
\begin{align}
T^{xx}_{i,i+1} &= \bra{\Psi_0}\sigma^x_i\sigma^x_{i+1}\ket{\Psi_0} = \langle {\cal B}_i{\cal A}_{i+1}\rangle \label{eqn: Txx}, \\
T^{yy}_{i,i+1} &= \bra{\Psi_0}\sigma^y_i\sigma^y_{i+1}\ket{\Psi_0} = - \langle {\cal A}_i{\cal B}_{i+1} \rangle \label{eqn: Tyy}, \\
T^{zz}_{i,i+1} &= \bra{\Psi_0}\sigma^z_i\sigma^z_{i+1}\ket{\Psi_0} = \langle {\cal A}_i{\cal B}_i{\cal A}_{i+1}{\cal B}_{i+1} \rangle \label{eqn: Tzz}, 
\end{align}
where $\langle {\cal B}_i{\cal A}_{i+1}\rangle = \bra{\Psi_0}{\cal B}_i{\cal A}_{i+1}\ket{\Psi_0}$, etc. Similarly, the off-diagonal correlations can also be obtained in terms of ${\cal A}_i$ and ${\cal B}_i$.\\

Using orthogonality of $\phi_k$ and $\psi_k$,  from Eqs. (\ref{eqn: eta}), (\ref{eqn: phi}), and (\ref{eqn: psi}), it follows that ${\cal A}_i = \sum_k (\eta_{k} + \eta_k^{\dagger})\phi_{ki}$ and ${\cal B}_i = \sum_k (\eta_{k} - \eta_k^{\dagger})\phi_{ki}$. Since ${\cal A}_i$'s and ${\cal B}_i$'s are anticommuting variables, their vacuum expectation values can be evaluated using Wick's theorem and we get
\begin{align}
\langle {\cal A}_i{\cal A}_j \rangle &= \sum_k \phi_{ki}\phi_{kj} = \delta_{ij} \label{eqn: AiAj},\\
\langle {\cal B}_i{\cal B}_j \rangle &= - \sum_k \psi_{ki}\psi_{kj} = - \delta_{ij} \label{eqn: BiBj},\\
\langle {\cal B}_i{\cal A}_j \rangle &= - \langle {\cal A}_j{\cal B}_i \rangle = - \sum_k \psi_{ki}\phi_{kj} \nonumber \\
&=  - (\boldsymbol{\psi}^T \boldsymbol{\phi})_{ij} = G_{ij} \label{eqn: BiAj},
\end{align}
where $\boldsymbol{\phi}$ and $\boldsymbol{\psi}$ are the matrices of $\phi_{ki}$ and $\psi_{ki}$ respectively, and $G$ is the correlation matrix.

Using Eqs. (\ref{eqn: AiAj}), (\ref{eqn: BiBj}), and (\ref{eqn: BiAj}), we obtain $m^z_i = -G_{ii}$ and $m^x_{i} = m^y_{i} =0$. The diagonal correlations are reduced to
\begin{align}
T^{xx}_{i,i+1} &= G_{i,i+1}, \label{eqn: Txxf}\\
T^{yy}_{i,i+1} &= -G_{i,i+1}, \label{eqn: Tyyf}\\
T^{zz}_{i,i+1} &= G_{i,i}G_{i+1,i+1} - G_{i,i+1}G_{i+1,i}, \label{eqn: Tzzf}
\end{align}
while all off-diagonal correlations vanish.

The above formalism leads to the single- and two-site nearest-neighbor density matrices of the ground states of the ordered as well as disordered $XY$ spin models as
\begin{align}
&\rho_i = \Tr_{\widehat{i}}{(\ket{\psi_0}\bra{\psi_0})} =\frac 12 [I + m^z_i\sigma_i^z], \\
&\rho_{i,i+1} = \Tr_{\widehat{i,i+1}}{(\ket{\psi_0}\bra{\psi_0})} = \frac{1}{4}\Big[I\otimes I+m^z_i(\sigma^z\otimes I)  \nonumber\\
&\hspace{2.5em} + (I \otimes \sigma^z) m^z_{i+1}+ \sum_{\alpha=x,y,z}T_{i,i+1}^{\alpha\alpha}(\sigma^{\alpha}\otimes\sigma^{\alpha})\Big].
\end{align}
Here $\Tr_{\widehat{i}}(\cdot)$ denotes the tracing out from the argument of all sites except $i$.  $\Tr_{\widehat{i,i+1}}$ is similarly defined.

\subsection{Thermal States: $T \ne 0$}

The technique discussed above 
for the ground state can now be extended to the thermal equilibrium state. At any finite temperature $T$, the canonical equilibrium state is given by 
\begin{equation}
\rho(\beta) = \frac{\mbox{exp}(- \beta H)}{\cal Z},
\label{eq:equilib}
\end{equation}
where \(\cal Z\) is the partition function, 
\[\cal Z = \Tr[\mbox{exp}(-\beta H)], \]
and \(\beta = \frac{1}{k_{B}T}\), with \(k_B\) being the Boltzmann constant. 
The elements of the correlation matrix in this case are given by 
\begin{align}
G_{ij}(\beta) = \langle {\cal B}_i{\cal A}_j \rangle_{\beta},
\end{align} 
where $\langle \cdot \rangle_{\beta}$ denotes an average over the canonical equilibrium state at temperature $T$. Thus
\begin{align}
G_{ij}(\beta) &= \sum_{k,k'}\psi_{ki}\phi_{k'j} \langle(\eta_k^{\dagger}-\eta_{k})(\eta_{k'}^{\dagger} + \eta_{k'}) \rangle_{\beta} \nonumber\\
&= \sum_{k}{\psi_{ki}\phi_{kj}\Big(\langle \eta_k^{\dagger}\eta_{k}\rangle_{\beta} - \langle \eta_{k}\eta_k^{\dagger}\rangle_{\beta} \Big).} \nonumber
\end{align}
From the Fermi-Dirac statistics, it follows that $\langle \eta_{k}\eta_k^{\dagger}\rangle_{\beta}=({e^{\beta\Lambda_k} + 1})^{-1}$. Using this, we have
\begin{align}
G_{ij}(\beta) &= - \sum_{k}\psi_{ki}\phi_{kj}\tanh{\left(\frac{\beta \Lambda_k}{2}\right)} \nonumber\\
&=-\Big(\boldsymbol{\psi}^T \tanh{\left(\frac{\beta \boldsymbol{\Lambda} }{2}\right)} \boldsymbol{\phi} \Big)_{ij} \label{eqn: thG},
\end{align}
where ${\mathbf \Lambda}$ represents the diagonal matrix of $\Lambda_k$.
Using Eq. (\ref{eqn: thG}), one can evaluate all the single and two particle observables and hence the single- and two-site density matrices,
similar to that for the ground state. 

\subsection{Ordered and Disordered Systems}
\subsubsection{The ordered quantum anisotropic $XY$ spin chain}
The ordered system corresponds to the case where $J_i$'s and $h_i$'s are site independent and its Hamiltonian is given by
\begin{align}
\tilde{H} = \frac {J}{4}  \sum_{i = 1}^{N} [(1+\gamma) \sigma_i^x \sigma_{i+1}^{x} + (1 - \gamma ) \sigma_i^y \sigma_{i+1}^{y}] - \frac {h}{2} \sum_{i=1}^N \sigma_i^z.
\label{eqn: XYO}
\end{align}
After the Jordan-Wigner, Fourier, and Bogoliubov transformations, the above Hamiltonian reduces to
%
%
\begin{align}
\tilde{H} &= 2 \sum_{k} \tilde{\Lambda}_k \tilde{\eta}_k^{\dagger} \tilde{\eta}_k - \sum_k \tilde{\Lambda}_k, \\
\intertext{where}
\tilde{\Lambda}_k &= \sqrt{(\gamma \lambda \sin{\phi_k})^2 + (1+ \lambda \cos{\psi_k})^2}, \nonumber
\end{align}
$\lambda = \frac Jh,$ and $\phi_k = 2\pi k/N; k = -\frac N2 , \ldots , 0,1, \ldots, \frac N2 - 1$.
Note that in the absence of disorder, all single-  and two-site nearest-neighbor density matrices are equivalent and are obtained by using magnetization and correlation functions.  Due to this translational symmetry, it is possible to calculate the eigenvalue spectrum, magnetization and correlation functions analytically \cite{barouch1, barouch2}.  \\

\subsubsection{Quantum $XY$ spin glass}

The Hamiltonian for the one dimensional quantum $XY$ spin glass is given by

\begin{align}
H^{SG}=\sum_{i}^N \frac{J_i}{4}[(1 + \gamma)\sigma_i^x\sigma_{i + 1}^x+(1 - \gamma)\sigma_i^y\sigma_{i + 1}^y] -\frac{h}{2}\sum_{i}^N\sigma_i^z, \label{eqn: XYDJ}
\end{align}
where the \(J_i\)'s are chosen as independent and identically distributed (i.i.d.) Gaussian random variables i.e., each follows the Gaussian (normal) distribution, $N(\langle J_i \rangle,\sigma)$ with $\langle J_i \rangle$ and $\sigma$ being the corresponding mean and standard deviation respectively. Here $h$ are assumed to be site independent and constant. We choose $\langle J_i \rangle$ to be independent of $i$ and set $\langle \lambda \rangle =  \frac{\langle J_i \rangle}{h}$.


\subsubsection{Random-field quantum $XY$ spin chain}

If the coupling constant is kept as site independent  and the randomness is introduced in the strength of the magnetic field,
the Hamiltonian for the corresponding quantum $XY$ spin chain with random transverse field is given by
\begin{align}
H^{RF}=\frac{J}{4}\sum_{i}^N  [(1 + \gamma)\sigma_i^x\sigma_{i + 1}^x+(1 - \gamma)\sigma_i^y\sigma_{i + 1}^y] -\sum_{i}^N\frac{h_i}{2}\sigma_i^z, \label{eqn: XYDh}
\end{align}
where the  \(h_i\)'s are i.i.d. Gaussian random variables, each following the Gaussian distribution with mean $\langle h_i \rangle$  and standard deviation $\sigma$. We choose $\langle h_i \rangle$ to be site independent and set $\langle \mu \rangle =  \frac{\langle h_i \rangle}{J}$.


\section{Enhancement Scores}
\label{sec:EnhancementScore}

As discussed in the preceding section, we consider disordered systems in which disorder is introduced either in the coupling strength or in the transverse magnetic field. To verify whether it is possible to observe disorder-induced-order (order-from-disorder) phenomena for some observable ${\cal Q}$, 
we introduce a quantity, called enhancement score for ${\cal Q}$ \cite{UtkarshCI}. At zero temperature, we define the enhancement score of the observable, ${\cal Q}$, as
%
 \begin{align}
  \Delta^{{\cal Q}}_{\lambda}&=|{{\cal{Q}}_{av}(\langle \lambda \rangle)|-|{\cal{Q}}(\langle\lambda\rangle)}| \nonumber\\ 
 \mbox{and} \hspace{1em} \Delta^{{\cal Q}}_{\mu}&=|{{\cal{Q}}_{av}(\langle \mu \rangle)|-|{\cal{Q}}(\langle\mu\rangle)}|, \label{eqn: GsAdv}
 \end{align}
where ${\cal{Q}}_{av}(\langle \lambda \rangle)$ represents the quenched averaged value of ${\cal Q}$ over the corresponding i.i.d. Gaussian random variables with mean $\langle \lambda \rangle$ and standard deviation $\sigma$ for the ground state. Similarly, one can define ${\cal{Q}}_{av}(\langle \mu \rangle)$. ${\cal{Q}}(\langle\lambda\rangle)$ and ${\cal{Q}}(\langle\mu\rangle)$ denote the corresponding quantities for the ordered system ground state with $\frac Jh = \langle \lambda \rangle$ and $\frac hJ=\langle \mu \rangle$ respectively.
$\Delta^{\cal Q}_{\lambda} > 0$ indicates the appearance of order-from-disorder phenomenon for ${\cal Q}$ in the system governed by the spin glass Hamiltonian. Similarly for $ \Delta^{{\cal Q}}_{\mu}$.

Thermal fluctuations, in general, destroy quantumness of the system and it behaves as a global disorder to the entire system. Similar to the spirit of the zero temperature enhancement score, we define the thermal enhancement scores of a physical quantity ${\cal Q}$ as 
%
\begin{align}
  \Delta^{{\cal Q}}_{\beta, \lambda}&={|{\cal{Q}}_{av}(\beta, \langle \lambda \rangle)|-|{\cal{Q}}(\beta,\langle\lambda\rangle)|}   \nonumber\\
 \mbox{and} \hspace{1em} \Delta^{{\cal Q}}_{\beta,\mu}&={|{\cal{Q}}_{av}(\beta,\langle \mu \rangle)|-|{\cal{Q}}(\beta,\langle\mu\rangle)|},  \label{eqn: TsAdv}
 \end{align}
where $\cal Q$ is measured in the thermal state $\rho(\beta)$ of the corresponding Hamiltonian at equilibrium temperature $T$. A positive value of   $\Delta^{{\cal Q}}_{\beta,\lambda}$ signals the order-from-disorder phenomenon in the thermal state of the spin glass system. Similarly for  $\Delta^{{\cal Q}}_{\beta,\mu}$.\\

Finally, we study whether it is possible to obtain an enhancement for an observable at a finite temperature that is better than the same at zero temperature, we introduce the total enhancement scores which are given by 
\begin{align}
  \Delta^{{\cal Q}}_{Total,\, \lambda}&= \Delta^{{\cal Q}}_{\beta,\lambda} - \max[0,\Delta^{{\cal Q}}_{\lambda}] \nonumber \\
 \mbox{and} \hspace{1em} \Delta^{{\cal Q}}_{Total,\, \mu}&= \Delta^{{\cal Q}}_{\beta, \mu} - \max[0,\Delta^{{\cal Q}}_{\mu}].  \label{eqn: TotalAdv}
 \end{align}
Positive values of $\Delta^{{\cal Q}}_{Total,\, \lambda}$ and  $\Delta^{{\cal Q}}_{Total,\, \mu}$ indicate the region where order-from-disorder phenomenon is more profound in the thermal state than in the zero-temperature states.

\section{Quantum Correlation Measures} 
\label{sec:qcmeasures}

Investigations on enhancement scores in the disordered systems are carried out by considering bipartite quantum correlation measures as well as classical correlations and magnetization.  Here we briefly describe the quantum correlation measures that we use in this paper. The thermal as well as the ground states of the Hamiltonians consist of $N$ spin-$\frac 12$ particles. To study the bipartite quantum correlations, we trace out all the particles except two nearest-neighbor ones. Since the system is with periodic boundary condition, all the nearest-neighbor density matrices are the same in case of the ordered system, while in the disordered case, the values of any observable corresponding to the nearest-neighbor states are again the same  after quenching. For the investigations, four quantum correlation measure are considered here, viz. concurrence and logarithmic negativity as entanglement measures, and quantum discord and quantum work-deficit as information-theoretic quantum correlation measures.


\subsection{Concurrence}

Let $\rho_{AB}$ be the density matrix corresponding to an arbitrary two-qubit system shared between two parties $A$ and $B$. The product, $\rho_{AB}\tilde{\rho}_{AB}$, though non-Hermitian, have only real and positive eigenvalues, say $ e_{1},\, e_{2},\,  e_{3},\, \mbox{and}\, e_{4}$ in descending order, where $\tilde{\rho}_{AB}=(\sigma_y \otimes \sigma_y) \rho^*_{AB} (\sigma_y \otimes \sigma_y)$, with $\rho^*_{AB}$ being the complex conjugate of the density matrix $\rho_{AB}$.
The concurrence  \cite{conc} for the two-qubit state $\rho_{AB}$ is defined as
\begin{equation}
 C(\rho_{AB})= \max [0,\sqrt{ e_{1}}-\sqrt{ e_{2}}-\sqrt{ e_{3}}-\sqrt{ e_{4}}].
\end{equation}


\subsection{Logarithmic Negativity}

Logarithmic negativity (LN)  \cite{logneg} is a computable measure of entanglement for any mixed state of an arbitrary bipartite system. LN is based on the definition of negativity \cite{logneg}, which is given by
\begin{equation}
N(\rho_{AB})=\frac{||\rho^{T_A}||-1}{2}.
\end{equation}
Here $||\rho_{AB}^{T_A}||$ is the trace norm of the partially transposed density matrix $\rho_{AB}^{T_A}$ with the partial transposition being taken on subsystem  $A$ \cite{partialtranspose}.  
The LN is defined as 
\begin{equation}
E_{N}=\textrm{log}_2||\rho^{T_A}||=\textrm{log}_2[2N(\rho_{AB})+1].
\end{equation}
LN is non-vanishing for all entangled states of two spin-$\frac 12$ particles \cite{partialtranspose} and can be used to quantify the degree of the entanglement in all composite systems.

%


\subsection{Quantum Discord}

Mutual information between classical random variables can be defined in two equivalent ways.
If $X$ and $Y$ are two random variables which assume values $x_i$ and $y_i$ with probabilities $p_i$ and $q_i$ respectively, then the total correlation between the variables $X$ and $Y$, quantified by the mutual information, is defined as
\begin{equation}
{\cal I}(X;Y) = H(X) + H(Y) - H(X,Y).
\label{eq:cmi1}
\end{equation}
Here $H(X)=-\sum_i p_i \log_2 p_i$ is the Shannon entropy for the random variable $X$, and similarly for $H(Y)$ and $H(X,Y)$. 
The second form of classical mutual information is defined using Bayesian rules as
\begin{eqnarray}
{\cal J}(X;Y)&=&H(X)-H(X|Y),
\label{eq:cmi2}
\end{eqnarray}
where $H(X|Y)=H(X,Y)-H(Y)$ is the conditional entropy. 
These two definitions are classically equivalent. However, the quantum analogs of Eqs. (\ref{eq:cmi1}) and (\ref{eq:cmi2}) are inequivalent and their difference is defined as quantum discord  \cite{discord}. The quantum versions for a bipartite quantum state $\rho_{AB}$ are given by
\begin{eqnarray}
I(\rho_{AB})&=&S(\rho_A)+S(\rho_B)-S(\rho_{AB}) \nonumber \\
\mbox{and}\,\,\,  J(\rho_{AB})&=&S(\rho_A)-S(\rho_{A|B}).
\end{eqnarray}
Here $S(\varrho)=-\Tr(\varrho \log_2 \varrho)$ and the quantum conditional entropy is given by $S(\rho_{A|B}) =  \min_{\{B_{i}\}} \sum_{i}p_{i}S(\rho_{A|i})$,
where the measurement is performed by $B$ with a rank-one projection-valued measurement, $\{B_i\}$, producing the ensemble $\{p_i,\, \rho_{A|i}=\frac{1}{p_i}\Tr_B[(\mathbb{I}_A\otimes B_i) \rho_{AB} (\mathbb{I}_A\otimes B_i)]\}$. Here $\mathbb{I}_A$ is the identity operator on the Hilbert space of $A$ and $p_i = \Tr[(\mathbb{I}_A\otimes B_i) \rho_{AB} (\mathbb{I}_A\otimes B_i)]$. The quantum discord is defined as 
\begin{equation}
D(\rho_{AB}) =I(\rho_{AB}) - J(\rho_{AB}),
\end{equation}
where $I(\rho_{AB})$ and $J(\rho_{AB})$ are respectively identified as total correlations and classical correlations in $\rho_{AB}$. 

\subsection{Quantum Work-Deficit}

For a bipartite quantum state, $\rho_{AB}$, quantum work-deficit  \cite{wd} is defined as the difference between the amount of pure states that can be obtained under global operations and pure product states that can be extracted under local operations, in closed systems for which addition of ancillary pure product states are not allowed. The number of pure qubits that can be extracted from a state $\rho_{AB}$ by
``closed global operations", which are sequences of unitary operations and dephasing operations, is given by 
\begin{equation}
I_G(\rho_{AB})=N-S(\rho_{AB}),
\end{equation}
where $N=\log_2 (\dim {\cal H}_{AB})$, with $ {\cal H}_{AB}$ being the Hilbert space on which $\rho_{AB}$ is defined. The  work extractable locally from $\rho_{AB}$ under ``closed local operations and classical communication" (CLOCC), which consists of local unitaries, local dephasing, and sending dephased state from one party to another, is defined as
\begin{equation}
I_L(\rho_{AB})=N- \min_{\{B_{i}\}} \sum_{i}S\left((\mathbb{I}_{A}\otimes B_{i})\rho_{AB}(\mathbb{I}_{A}\otimes B_{i})\right).
\end{equation}
Here the measurement is performed by $B$ with a rank-one projection-valued measurement $\{B_i\}$. The quantum work-deficit is given by
\begin{equation}
W(\rho_{AB})=I_G(\rho_{AB})-I_L(\rho_{AB}).
\end{equation}

\begin{figure*}[htb]
\includegraphics[angle=0,width=18cm]{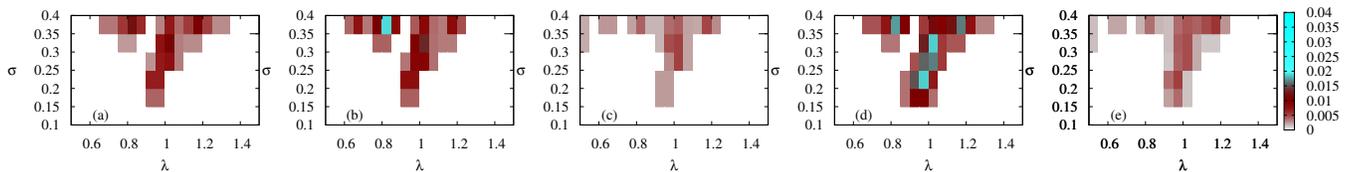}
\caption{(Color online.) Density plot of  difference between self-averaged and quenched averaged observables, i.e., of ${\cal S}^{\cal Q} = \frac 1N \sum_{i=1}^N {\cal Q}(\rho_{i,i+1}) - \int \int \cdots \int {\cal Q}(\{\lambda_i\}) d\{\lambda_i\} $  against $\lambda$ on the horizontal axis,  and the disorder strength, $\sigma$ on the vertical axis for the quantum $XY$ spin glass Hamiltonian. Here $\lambda_i=\frac{J_i}{h}$. We choose $N=10^3$ and  $\gamma = 0.7$. From left panel to right,  the observables are chosen as  (a) magnetization, (b) $C_{xx}$, (c) $C_{yy}$, (d) $C_{zz}$, and (e) concurrence.  White region indicates ${\cal S}^{\cal Q}=0$ which implies that the self-averaged and quenched averaged observables are equivalent. Here, maximal error is of order 0.005. 
}
\label{fig: saclscl}
\end{figure*}

\subsection{Quenched Averaging vs. Self Averaging}
\label{subsec:quench}

The disordered physical parameters of the disordered systems studied here are considered to be ``quenched", i.e., the time scale in which the dynamics of the system takes place is much shorter than the equilibrating time of the disorder. 
So, 
the averaging over the random variables has to be performed after the calculation of the physical quantities, for a given state. Specifically, the quenched averaged value for a physical quantity, ${\cal Q}$, is defined as 
\begin{equation}
{\cal Q}_{av}(\langle\lambda\rangle)= \int \int \cdots \int {\cal Q}(\{\lambda_i\}) d\{\lambda_i\},
\end{equation}
where the integration is performed over the Gaussian distributions $\{\lambda_i\}$, and  ${\cal Q}(\{\lambda_i\})$ is the value of the observable ${\cal Q}$ in the state under study (say, the ground state) of the with system parameters $\lambda_i$.

The entire analyses for disordered as well as ordered systems are carried out by considering the system to be of $10^3$ quantum spin-$\frac 12$ particles arranged on a chain with periodic boundary conditions, and we are interested to study the observables after quenching. In case of $10^3$ quantum spin-$\frac 12$ particles interacting according to a disordered Hamiltonian, convergence of an observable after the quench to a fixed value requires the convergence of an integration in a space of dimension of about $10^3$. To overcome such computational difficulties, we check whether the observables like magnetization, correlations, and bipartite quantum correlation measures are self-averaging quantities. A two-site physical quantity ${\cal Q}$ is said to be self-averaging if \cite{selfavebook}
\begin{align}
{\cal Q}_{av}(\langle\lambda\rangle)\equiv\int \int \cdots \int {\cal Q}(\{J_i\}) d\{J_i\} = \frac 1N \sum_{i=1}^N {\cal Q}(\rho_{i,i+1}).
\end{align}
We find that all the physical quantities that we use in our analysis, self average, when the system is far from the critical point (see Fig. \ref{fig: saclscl}).


Note that when $\sigma \rightarrow 0$, the system becomes  the ordered $XY$ model and so the question of self-averaging is irrelevant in that limit. 
The ground state of the ordered $XY$ model undergoes a quantum phase transition at $\frac Jh = 1$, and one can check that site-averaging and sample-averaging are not equivalent in the vicinity of the quantum critical point.
It becomes prominent with the increase of $\sigma$ (see Fig. \ref{fig: saclscl}).
Throughout the paper, we set a moderate value for $\sigma$, viz., $\sigma=0.3$, where in we observe that except when $\langle\lambda\rangle\approx 1$ or when $\langle\mu\rangle\approx 1$ all the observables self average. For simplicity, we will henceforth denote $\langle\lambda\rangle$ as $\lambda$ and $\langle\mu\rangle$ as $\mu$, even for the disorder systems.

\section{Order-from-disorder: Quenched disorder at finite temperature}
\label{sec:mainresults}

In this section, we investigate the enhancement of different physical observables due to disorder in coupling as well as field strengths in the one dimensional transverse quantum $XY$ model. Specifically, we study the enhancement scores of different quantum correlation measures in the presence of disorder. The two subsections deal respectively with the cases of quenched disorders in the coupling strengths and in the transverse magnetic field. In both cases, the effect of finite temperature is also analyzed. We see that logarithmic negativity and concurrence behave in a similar fashion. Similarly, the behaviors of quantum discord and quantum work-deficit are similar. This is both at finite as well as zero temperatures. Without loss of generality, all the plots given in this paper are for concurrence and quantum discord. 

\subsection{Disorder in Coupling Constant: Spin Glass}

Let us first study the behavior of the anisotropic $XY$ spin glass model in a transverse magnetic field.
We begin with the ground state, and then go over to finite temperatures.

\subsubsection{Ground state enhancement} 

\begin{figure}[t]
\includegraphics[angle=0,width=8cm]{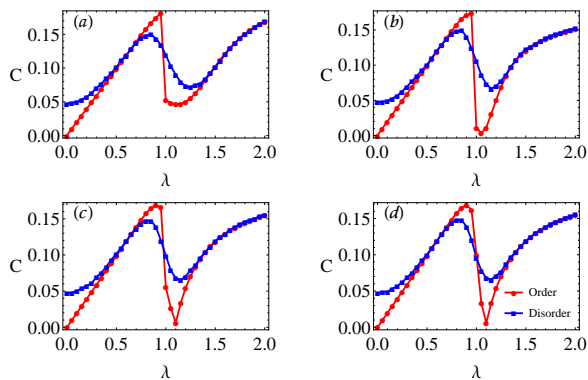}
\caption{(Color online.) Order-from-disorder for concurrence. Comparison between the concurrences of the nearest-neighbor density matrices of the zero-temperature states of the ordered $XY$ model  (red circles) and that of the $XY$ spin glass  system (blue squares)  against $\lambda$ for different system sizes. Here, $\gamma =0.4$ and $\sigma=0.3$. The top panels, (a) and (b) are for $N=8$ and $N=12$ respectively while the bottom panels, (c) and (d) represent $N=20$ and $N=100$ respectively. The horizontal axis is dimensionless, while the vertical one is in ebits. Note that the horizontal axis represents $\frac Jh$ for the ordered system curve, while the same represents $\frac{\langle J\rangle}{h}$ for the disordered one.}
\label{fig: GsScalingC}
\end{figure}

\begin{figure}[t]
\includegraphics[angle=0,width=8cm]{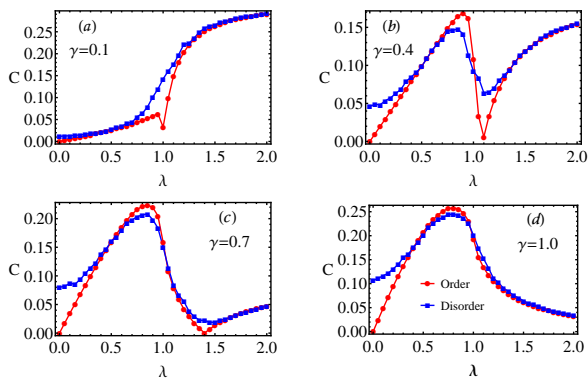}
\caption{(Color online.) Order-from-disorder for concurrence for different $\gamma$ and for $N=10^3$. All other considerations remain same as in Fig. \ref{fig: GsScalingC}. }

\label{fig: CGsJ}
\end{figure}

We compare the behavior of all the quantum correlation measures, defined in Sec.\ \ref{sec:qcmeasures}, of the ground state of the disordered $XY$ spin glass model, for different anisotropy parameters, with those in the ordered ones. At $\lambda = 0$, all the entanglement measures and information theoretic ones of the ordered system vanish while it is not the case for the disordered system and hence $\Delta^Q_{\lambda} > 0$ at $\lambda = 0$.  However, there are finite regions of the $\lambda$-axis, including those not containing $\lambda =0$, which exhibit $\Delta_{\lambda}^{\cal Q}>0$. In particular, $\Delta_{\lambda}^C>0$ when $0 \le \lambda \lesssim 0.55 $, as well as when $1 \le \lambda \lesssim 1.70$ with $\sigma = 0.3$, and  $\gamma = 0.4$. See Figs. \ref{fig: GsScalingC} and \ref{fig: CGsJ}. 
From Fig.\ \ref{fig: CGsJ}, one can notice that entanglement of the ordered $XY$ model shows sudden collapse and revival with respect to $\frac{J}{h}$ for moderate values of $\gamma$, which do not occur in the disordered model. 
Moreover, with the increase of $\gamma$, the region of positive enhancement score for concurrence decreases. In Fig.\ \ref{fig: CGsJ}, the investigation is carried out for $N=10^3$. It is interesting to study the trend of the enhancement region with the increase of $N$. Fig.\ \ref{fig: GsScalingC} shows this behavior for concurrence, with different N. 
It is clear that the region for which $\Delta_{\lambda}^C>0$  converges to a fixed region (up to the third decimal point) even for a relatively small system size, like $N=20$. Due to such convergence obtained here as well as for other quantum correlation measures and for other spin models, the results presented here for the quenched disordered quantum spin models are true in the thermodynamic limit.


\begin{figure}[t]
\includegraphics[angle=0,width=8cm]{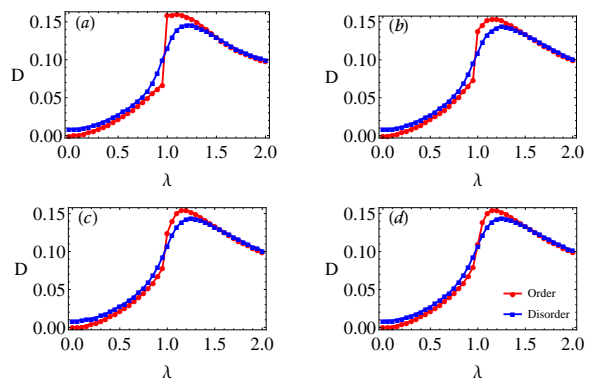}
\caption{(Color online.) Order-from-disorder for quantum discord. The vertical axis is measured in bits. All other considerations remain the same as in Fig. \ref{fig: GsScalingC}.}
\label{fig: GsScalingD}
\end{figure}

\begin{figure}[t]
\includegraphics[angle=0,width=8cm]{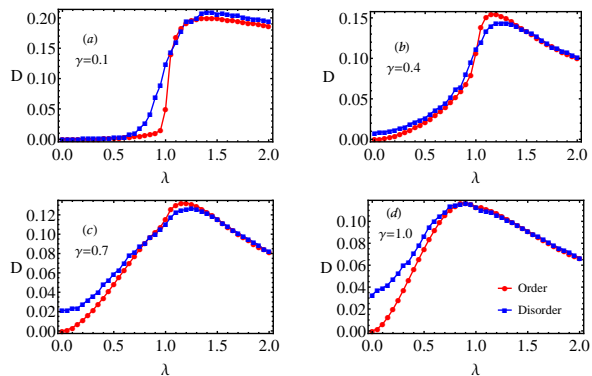}
\caption{(Color online.) Order-from-disorder for quantum discord for different $\gamma$ and for $N=10^3$. The vertical axis is measured in bits. All other considerations remain the same as in the Fig. \ref{fig: GsScalingC}.} 
\label{fig: DGsJ}
\end{figure}

In a similar spirit, we investigate the behavior of the discord enhancement score with the increase of $\lambda$. The information-theoretic measures (quantum discord and quantum work-deficit)  behave in a qualitatively different  way than that of the entanglement measures. The plots of the enhancement score for quantum discord are given in Figs. \ref{fig: GsScalingD} and \ref{fig: DGsJ}. Comparing Figs.\ \ref{fig: CGsJ} and \ref{fig: DGsJ}, we observe that instead of two regions, the discord enhancement score is positive for only one region, but the range of $\lambda$ in which disordered system possesses higher values gets increased. For example, $\Delta^D_{\lambda}>0$  in the region $0\le \lambda \le 1$ for $\gamma = 0.4, \sigma =0.3$ (see Fig.\ \ref{fig: DGsJ} for different values of $\gamma$). The scaling of the region with $\Delta^D_{\lambda}>0$ is shown in Fig.\ \ref{fig: GsScalingD} and the exact numerical values are listed in Table \ref{tab: disc_gs_scaling}.

\begin{table}[h]
\begin{center}
\begin{tabular}{ l | c  } 
\hline 
N &  $\Delta^D_{\lambda}>0$  region  \\ \hline 
6 & 0.0 - 0.97  \\
8 & 0.0 - 0.97  \\
12 & 0.0 - 0.97  \\
16 & 0.0 - 0.97  \\
20 & 0.0 - 0.97  \\
50 & 0.0 - 0.98 \\
100 & 0.0 - 0.99 \\ \hline 
\end{tabular}
\end{center}
\caption{The regions with $\Delta^D_{\lambda}>0$, for $\gamma=0.4,\, \sigma=0.3$, and where the quantum discords are correct up to the third decimal point.}
\label{tab: disc_gs_scaling}
\end{table}

\subsubsection{Thermal enhancement score: Nonmonotonicity  with temperature} 

\begin{figure}[h!]
\includegraphics[angle=0,width=8.25cm]{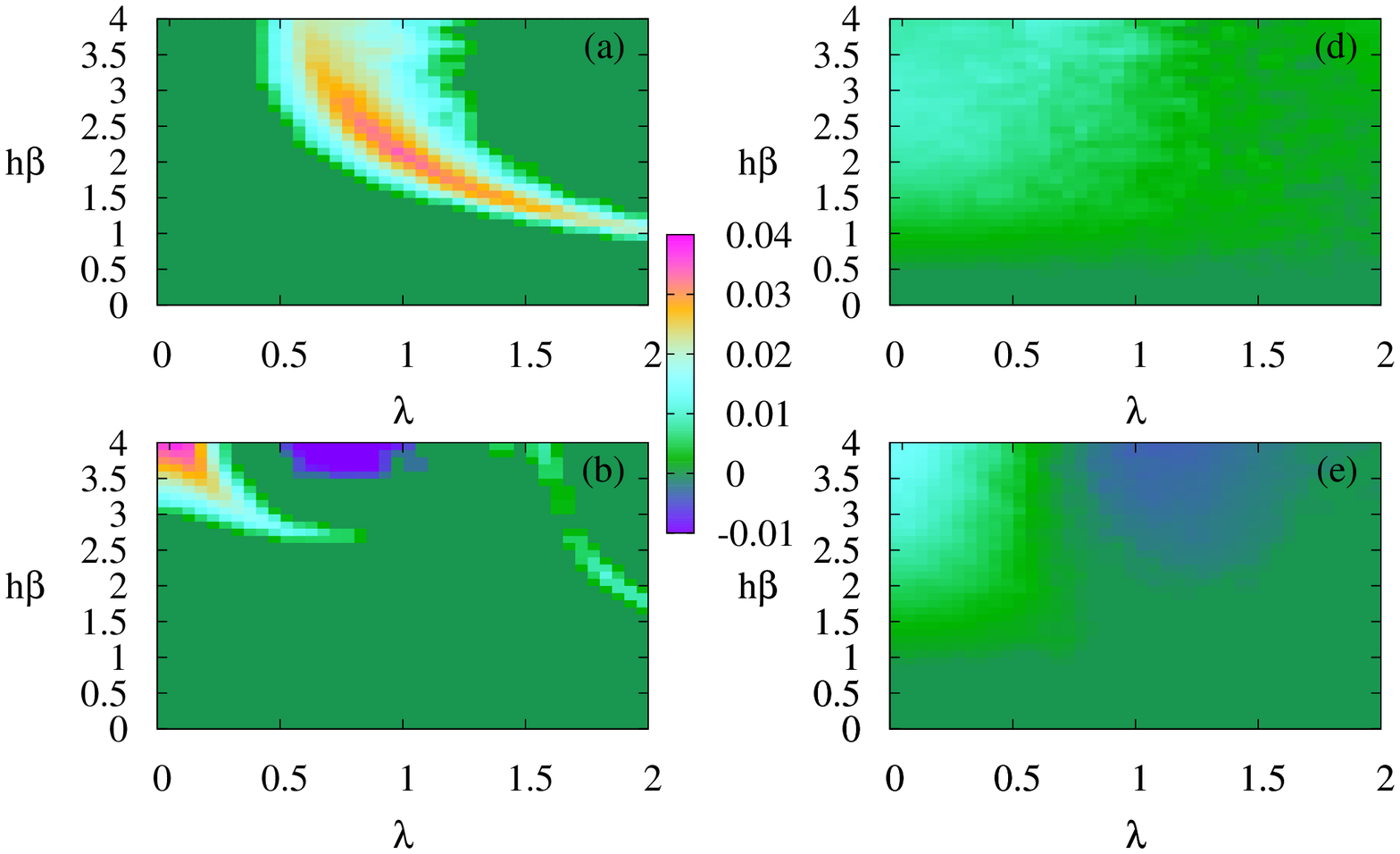}\\
\includegraphics[angle=0,width=4cm]{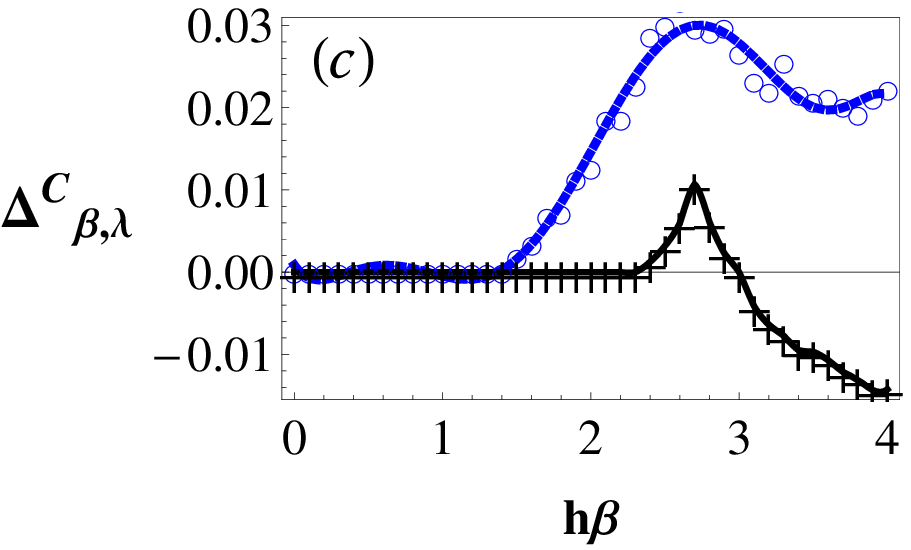}
\includegraphics[angle=0,width=4cm]{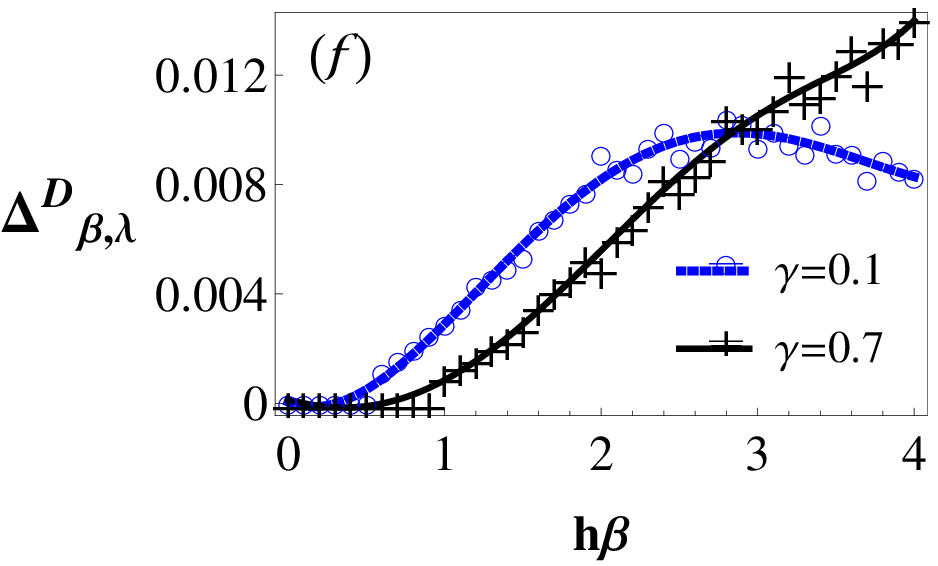}
\caption{(Color online.) Left panels, (a) and (b): Density plot of thermal enhancement scores for concurrence against $\lambda$ as abscissa and $h\beta$ as ordinate, for  (a) $\gamma =0.1$ and (b) $\gamma=0.7$, for the $XY$ spin glass model. We set $N=10^3$ and $\sigma =0.3$. Left bottom panel, (c): $\Delta^C_{\beta,\lambda}$  vs. $h\beta$ for fixed $\lambda=0.8$. Other parameters are the same as in the top panels. Clearly it shows nonmonotonicity of the entanglement enhancement scores with temperature for the $XY$ spin glass model. The right panels are for quantum discord. All other considerations are same in panels (d) and (e) as those in panels (a) and (b) respectively. Right bottom panel, (f):  $\Delta^D_\beta$  vs. $h\beta$ for $\lambda=0.1$.  Nonmonotonicity of the enhancement score for quantum discord is observed only for lower values of $\gamma$. The horizontal axes of all the panels are dimensionless. The vertical ares of the panels (a), (b), (d), and (e) are also dimensionless. $\Delta^C_{\beta,\lambda}$ is measured in ebits while $\Delta^D_{\beta,\lambda}$ is measures in bits.
}
\label{fig: CTsJ}
\end{figure}

Thermal fluctuations can effectively be viewed as global disorder introduced in the system, and hence one may expect that any order-from-disorder phenomenon observed for a quantum correlation measure could be less pronounced in the presence of a finite temperature, as compared to their values at zero temperature. 
On the contrary, we find that the thermal enhancement scores can be nonmonotonic with respect to temperature (see the bottom panels of Fig. \ref{fig: CTsJ}) (cf. \cite{nonmonotonicity}).
Moreover, there exists a region on the $(h\beta, \lambda)$-plane in which the thermal enhancement score is positive irrespective of quantum correlation measure used. In particular, this implies that for a fixed temperature, quantum correlation can be enhanced by putting disorder in the system.
Hence the ``order-from-disorder'' phenomena for quantum correlation can be observed also for moderate values of temperature. For example, 
we find that for $\lambda =1.5$ and $\gamma=0.4$, positive thermal enhancement scores, both for entanglement as well as information-theoretic measures, can be seen even for relatively high values of temperature (Fig.~\ref{fig: CTsJ}), and hence the increase of quantum correlation due to disorder in coupling at finite temperature can not be explained by the continuity argument of the same observation in the zero-temperature state. If one compares entanglement measures with the information-theoretic ones, it is evident that for a fixed temperature, entanglement enhancement scores possess much higher values compared to the information-theoretic ones, 
irrespective of the anisotropy parameter. Moreover, we observe that with the increase of $\gamma$, the regions with positive enhancement scores of entanglement measures shift from higher values of $\lambda$ towards $\lambda=0$, and at the same time, the area in which positive enhancement occurs gets reduced. In contrast, the information-theoretic measures always have a positive enhancement score near the $\lambda=0$ line.

\subsubsection{Total enhancement score}

\begin{figure}[h!]
\includegraphics[width=0.8\columnwidth,keepaspectratio,angle=0]{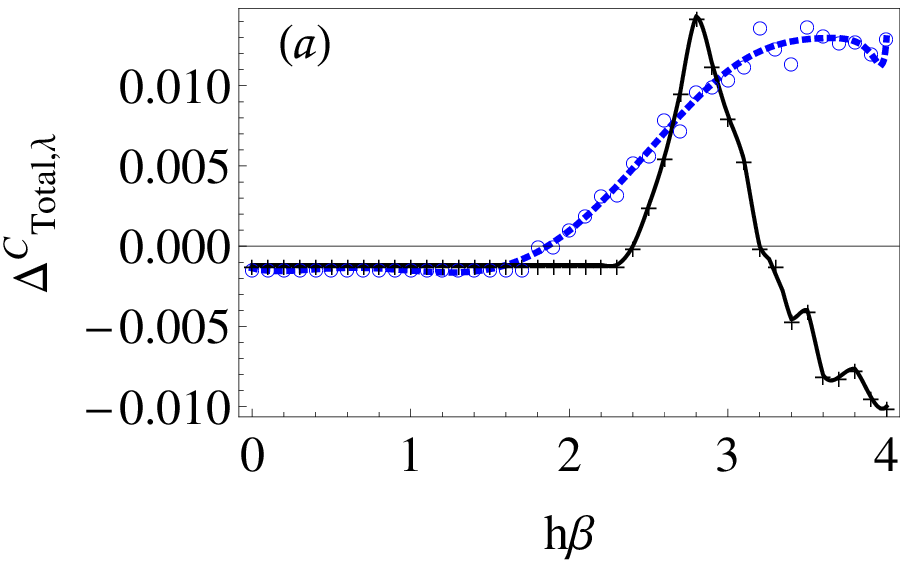}\\
\includegraphics[width=0.8\columnwidth,keepaspectratio,angle=0]{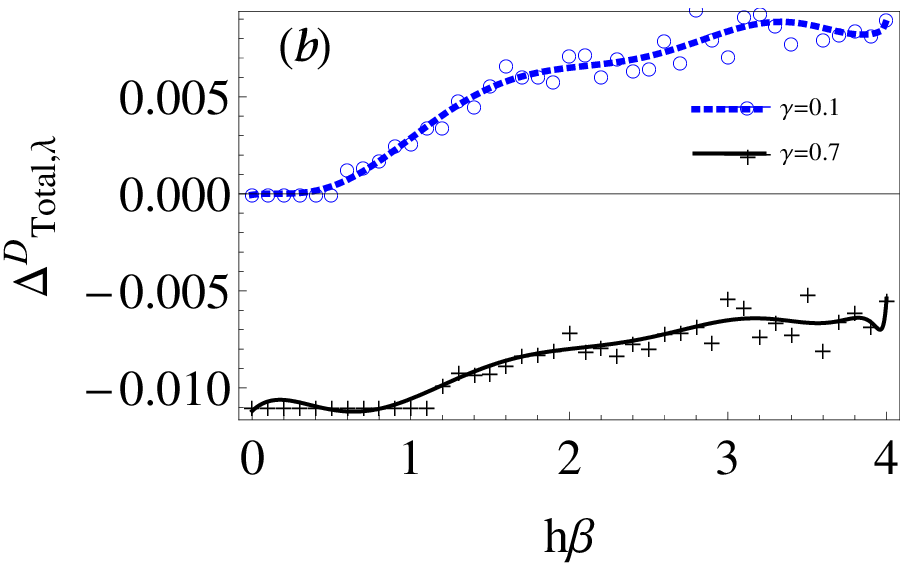}
\caption{(Color online.) Behavior of total enhancement score for concurrence (upper panel) and quantum discord (lower panel) for the thermal state 
of the quantum $XY$ spin glass. The data is fitted with the best polynomial fit. Two different values of anisotropy constants are chosen: $\gamma =0.1\, \textrm{and}\, 0.7$. Here, $\lambda=0.5$, $\sigma=0.3$, and $N=10^3$. The horizontal axes in both panels are dimensionless, while the vertical axis in the top (bottom) panel is in ebits (bits). }
\label{fig: CDTotalJ}
\end{figure}

The intuitive feeling of the fragility of entanglement and other quantum correlation measures leads us to believe that quantum correlations would decrease with temperature. We have already seen that this intuition is false in our discussions of the thermal enhancement score. To analyze this link further, we consider the total enhancement score, as defined in Sec. \ref{sec:EnhancementScore}.
When $\lambda$ is in the vicinity of  zero, all quantum correlation measures of the zero-temperature state show higher values of enhancement scores than that of the thermal state.
However, we find several values of $\lambda$ for which $ \Delta^{{\cal Q}}_{Total, \lambda}>0$ in a region on the $h\beta$-axis, for different values of $\gamma$ (see Fig. \ref{fig: CDTotalJ}). This finding is independent of the choice of the quantum correlation measure. For fixed $\gamma$ and $\lambda$, say $\gamma=0.1$ and $\lambda=0.5$, we observe that the total enhancement score for entanglement is positive for $h\beta \gtrsim 2$, while the same occurs for quantum discord for $h\beta \gtrsim 0.75$. However, the value of the total enhancement scores for discord is less than that for concurrence.

\subsection{Disorder in Transverse Field Strength: Random-field $XY$ model}

We now discuss the disorder-induced effects on quantum correlations when the disorder is introduced in the transverse field of the $XY$ model. One aim is to compare the behavior of quantum correlation in this case with that in the $XY$ spin glass. Since in this case, the randomness is introduced in the local part of the Hamiltonian, one may expect that the effects of disorder on quantum correlations due to randomness will be much less pronounced than for the $XY$ spin glass model.

\subsubsection{Enhancement scores: Zero and non-zero temperatures} 

\begin{figure}[h!]
\includegraphics[angle=0,width=8cm]{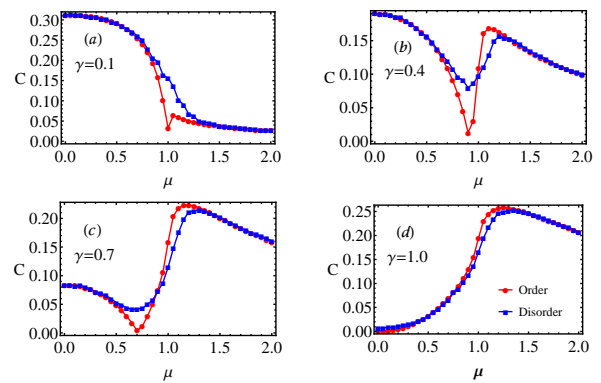}
\caption{(Color online.) Order-from-disorder for concurrence in the random-field $XY$model in the zero-temperature state. The red circles correspond to the ordered system, while the blue squares correspond to the disordered one. 
Here, 
$\sigma=0.3$ and $N=10^3$. The four panels are for different values of $\gamma$. The horizontal axes are dimensionless, while the vertical axes are in ebits.}
\label{fig: CGsh}
\end{figure}

\begin{figure}[h!]
\includegraphics[angle=0,width=8cm]{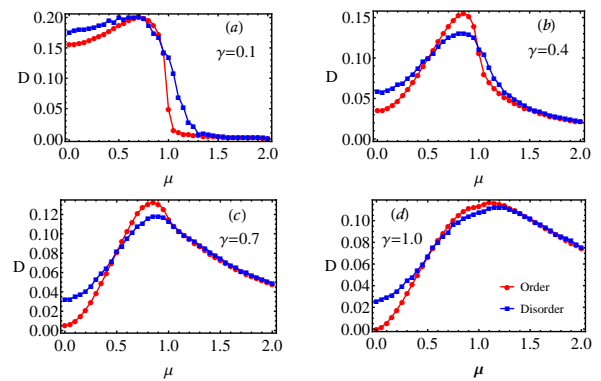}
\caption{(Color online.) Order-from-disorder for quantum discord. The vertical axes are in bits. All other considerations remain the same as in Fig. \ref{fig: CGsh}.}
\label{fig: DGsh}
\end{figure}

We begin with the behavior of entanglement with $\mu = \frac{\langle h \rangle}{J}$ at zero temperature. For the spin glass system, the entanglement enhancement scores were positive in two regions, while those for information-theoretic measures were positive in a single region. The situation exactly reverses in the random-field $XY$ model. See Figs. \ref{fig: CGsh} and \ref{fig: DGsh}. 

\begin{figure}
\includegraphics[angle=0,width=8.5 cm]{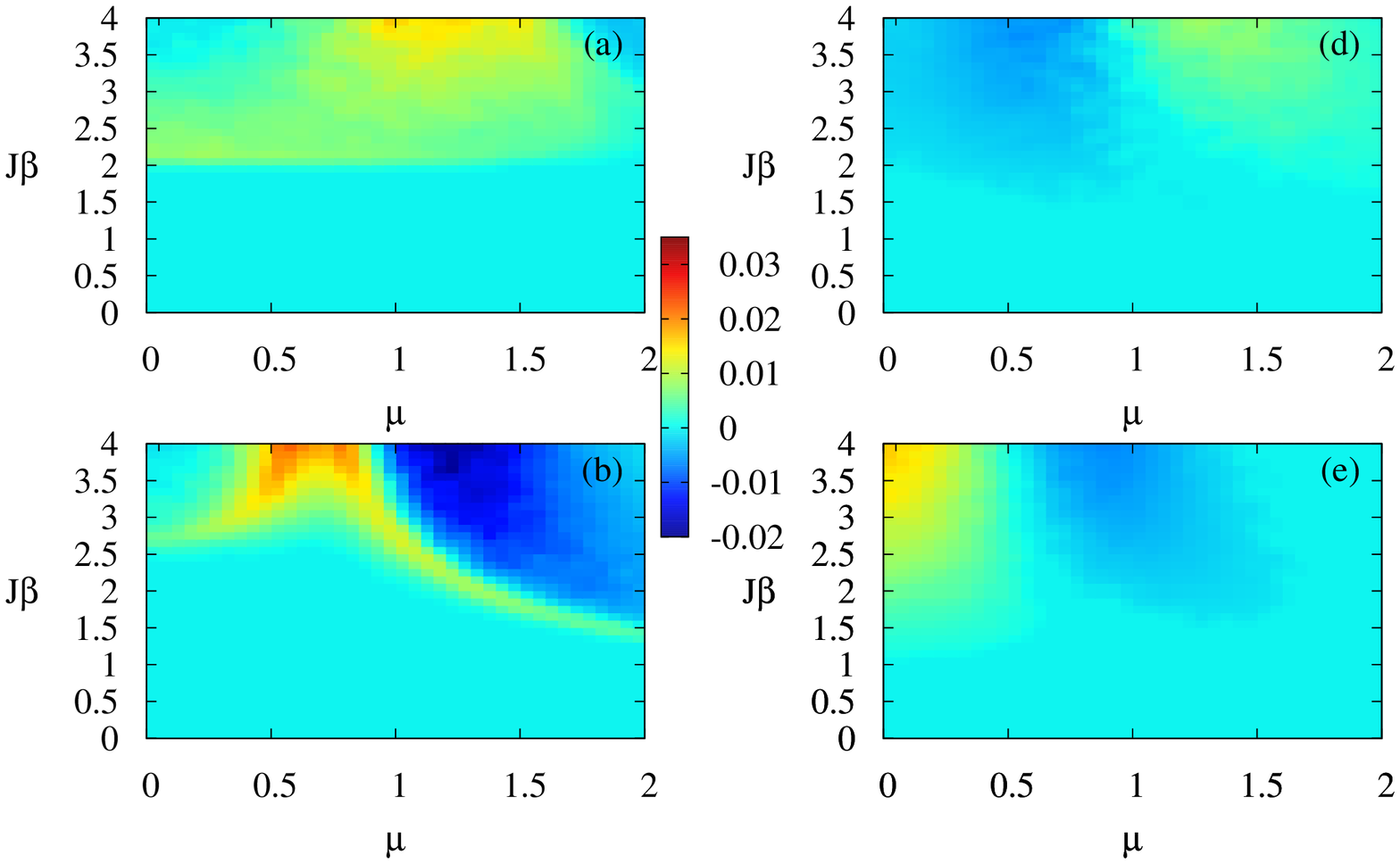}\\
\includegraphics[angle=0,width=4.0cm]{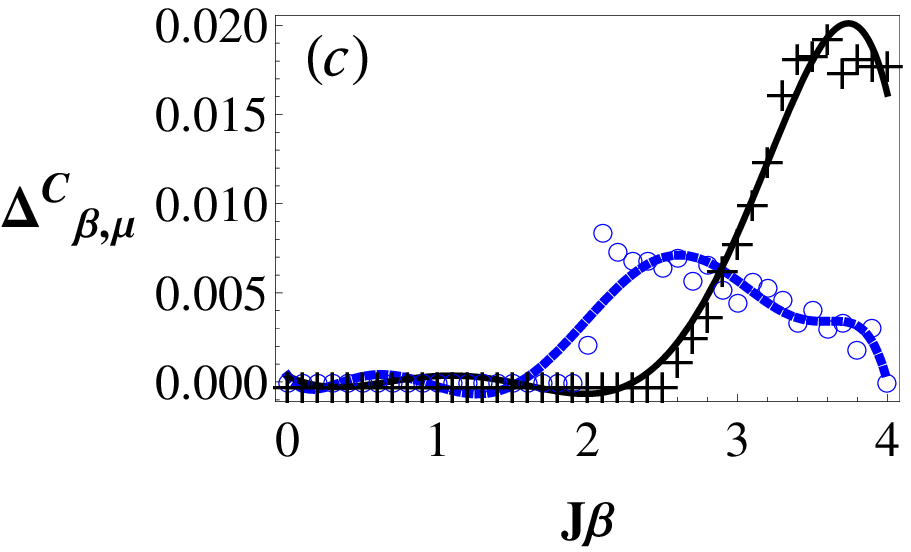}
\includegraphics[angle=0,width=4.0cm]{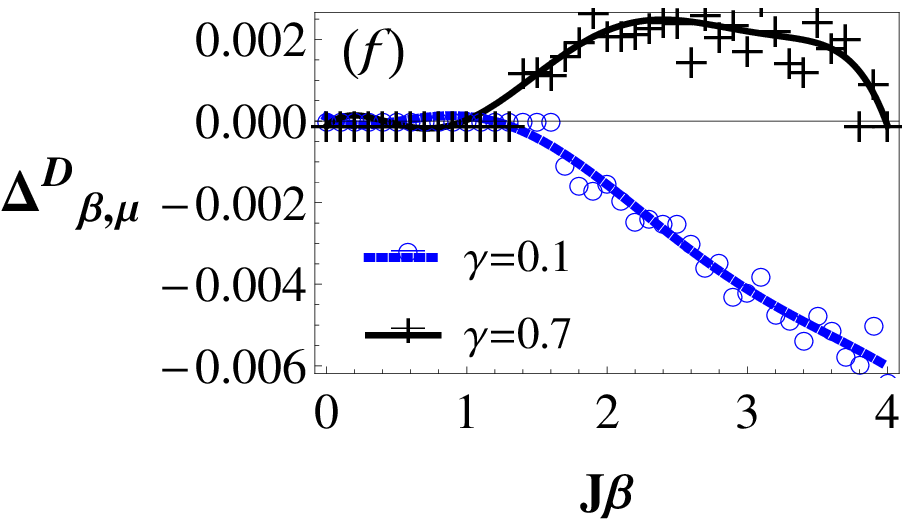}
\caption{(Color online.) The panels here are the same as in Fig. \ref{fig: CTsJ}, except that here they are for the transverse-field $XY$ model. Except for the simple changes for the change in the model, we have chosen $\mu=0.5$ in the two bottom panels.
 }
\label{fig: CTsh}
\end{figure}

Let us now consider the behavior of quantum correlation in the presence of both thermal fluctuation and randomness in the transverse field. The comparison is made between the thermal state of random-field  $XY$ model and that of the ordered one. Like in the case of the spin glass model, we find that the order-from-disorder behavior persist against thermal fluctuations for both concurrence and quantum discord.  See Fig. \ref{fig: CTsh}. Note that with the increase of $\gamma$, the positive enhancement score for concurrence near the $\mu=0$ line disappears, while the opposite is seen for quantum discord (see Fig. \ref{fig: CTsh}). Notice that especially for entanglement, the enhancement scores are more robust with the increase of temperature in the case of the $XY$ spin glass than  in the random-field $XY$ model.




\subsubsection{Total enhancement score}

\begin{figure}[t]
\includegraphics[width=0.85\columnwidth,keepaspectratio,angle=0]{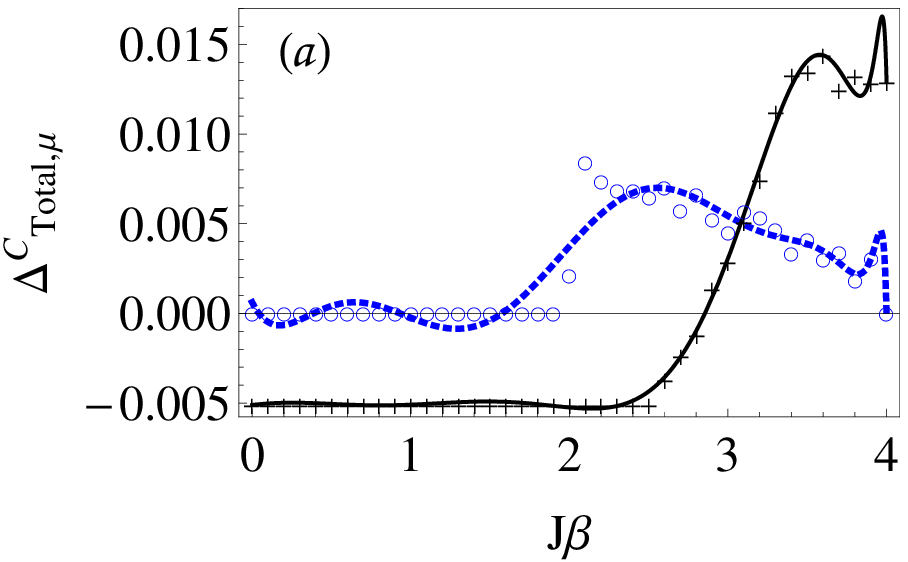}\\
\includegraphics[width=0.85\columnwidth,keepaspectratio,angle=0]{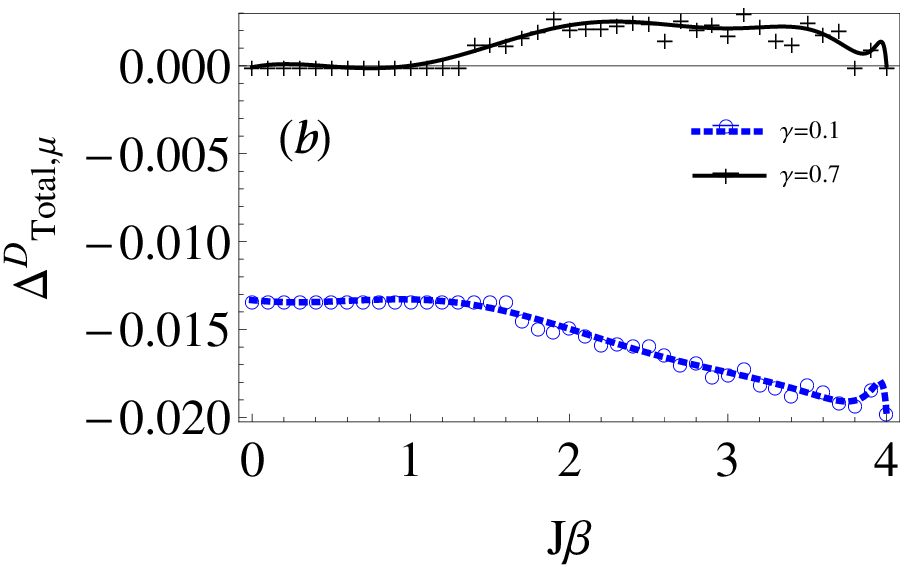}
\caption{(Color online.) The panels here are the same as in Fig. \ref{fig: CDTotalJ}, except that here they are for random-field $XY$ model. We choose $\mu=0.5$.}
\label{fig: CDTotalh}
\end{figure}

Just like the spin glass model, we again find that the total enhancement score for entanglement measures clearly possess positive value for moderate values of $J\beta$. For example, for $\gamma = 0.1$ and $\mu =  0.5$, the total enhancement score for concurrence is positive for $J\beta \gtrsim 2$, as depicted in Fig. \ref{fig: CDTotalh} (upper panel). This finding is independent of the choice of $\gamma$. In contrast, the total enhancement scores of information-theoretic measures, like quantum discord, does not show such clear signature of positivity in the presence of non-zero temperatures (see Fig. \ref{fig: CDTotalh} (lower panel)). 


As depicted in Fig.\ \ref{fig: CDTotalh} (upper panel), the total enhancement score for concurrence is positive for all  moderate values of $\gamma$. However, for quantum discord, near $\gamma = 0$, for example, for $\gamma=0.1$, we do not find a region in which  $\Delta^{D}_{Total,\mu}>0$. For higher values of $\gamma$, information-theoretic measures show a region on the $\mu$-axis, for which  $\Delta^{D}_{Total,\mu}>0$, although the value of $\Delta^{D}_{Total,\mu}$ is relatively small.

Comparing Figs. \ref{fig: CDTotalJ} and \ref{fig: CDTotalh}, we observe that there is a role reversal in the behavior of the total enhancement scores for both concurrence as well as for quantum discord, as we go over from higher to lower values of the anisotropy parameter and when we change the system from spin glass to the random-field $XY$ model. More specifically, the high-$\gamma$ (low-$\gamma$) spin glass model behave as the low-$\gamma$ (high-$\gamma$) random-field $XY$ model.

\section{Conclusion}
\label{sec:conclusion}

We have considered the disorder-induced effects on quantum correlations in two paradigmatic disordered quantum spin models, viz., the one-dimensional quantum $XY$ spin glass and random transverse field quantum $XY$ models. The disorders are assumed to be quenched. And the quantum correlations considered are chosen both from the entanglement-separability paradigm as well as from the information-theoretic one. We find that the systems support the order-from-disorder phenomenon for all the quantum correlations considered at both zero and finite temperatures. We utilize the concept of enhancement scores to quantify the phenomena. We find that the scores can actually get enhanced with the introduction of thermal fluctuations. Furthermore, we identify regions in the parameter space, in the post-quenched regime, where self-averaging of the quantum correlations occur, and where the same is absent, and find that the regions are related to the existence of a quantum critical point of the corresponding ordered system.

Temperatures close to absolute zero are difficult to achieve in the laboratories. And therefore, it is important to uncover whether a phenomenon remains robust with the application of thermal fluctuations. The results of this paper show that moderate temperatures can actually be a better candidate, than zero temperature, for observing the order-from-disorder phenomenon for a broad spectrum of quantum correlations.

The calculations were carried out for two entanglement measures, viz. concurrence and logarithmic negativity, and for two information-theoretic quantum correlation measures, viz. quantum discord and quantum work-deficit. The discussions in the paper are however mainly centered around the effects seen for concurrence and quantum discord, as those for logarithmic negativity and quantum work-deficit are broadly similar to their team-mates in the respective camps. 

The numbers corresponding to the phenomena reported are seen to have already converged for about 20 (or lower) quantum spins, while the calculations are also carried out for up to \(10^3\) spins, from which we wish to claim that the phenomena can be observed for finite systems as well as for systems in the thermodynamic limit.

\begin{acknowledgments}
R.P. acknowledges support from the Department of Science and Technology, Government of India, in the form of an INSPIRE faculty scheme at the Harish-Chandra Research Institute (HRI), India. We acknowledge computations performed at the cluster computing facility in HRI.
\end{acknowledgments}

\end{document}